\RequirePackage{lineno}

\documentclass[aps,prc,reprint,floatfix,showpacs,showkeys,superscriptaddress]{revtex4-1}
\usepackage{epsf}
\usepackage{amsmath,amssymb}
\usepackage{graphicx}
\usepackage{wrapfig}
\usepackage{longtable}
\usepackage{array}
\usepackage{color}

\newcommand{\etal}{{\it et al.}}
\def\deg{^\circ}

\begin{document}

\title{Measurements of the Separated Longitudinal Structure
Function $F_L$ from Hydrogen and Deuterium Targets at Low $Q^2$}

\author{V. Tvaskis}
\affiliation{Hampton University, Hampton, Virginia 23668}
\author{A. Tvaskis}
\affiliation{Hampton University, Hampton, Virginia 23668}
\author{I. Niculescu}
\affiliation{James Madison University, Harrisonburg, Virginia 22807}
\author{D. Abbott}
\affiliation{Thomas Jefferson National Accelerator Facility, Newport News, Virginia 23606}
\author{G.S.~Adams}
\affiliation{Rensselaer Polytechnic Institute, Troy, New York 12180}
\author{A.~Afanasev}
\affiliation{Thomas Jefferson National Accelerator Facility, Newport News, Virginia 23606}
\author{A.~Ahmidouch}
\affiliation{North Carolina A \& T State University, Greensboro, North Carolina 27411}
\author{T.~Angelescu}
\affiliation{Bucharest University, Bucharest, Romania}
\author{J.~Arrington}
\affiliation{Physics Division, Argonne National Laboratory, Argonne, Illinois 60439}
\author{R.~Asaturyan}
\affiliation{Yerevan Physics Institute, Yerevan, Armenia}
% ...

\date{\today}
\begin{abstract}
Structure functions, as measured in lepton--nucleon scattering,
have proven to be very useful in studying the quark dynamics within the nucleon. However, it is experimentally difficult to separately determine the longitudinal and transverse structure functions, and consequently
there are substantially less data available for the longitudinal structure function in particular. Here we present separated structure functions for hydrogen and deuterium at low four--momentum transfer squared, $Q^2 <$1~GeV$^2$, and compare these with parton
 distribution parameterizations and a $k_T$ factorization approach.
While differences are found, the parameterizations generally agree with the data even at the very low $Q^2$ scale of the data. The deuterium data show a smaller longitudinal structure function, and smaller ratio of longitudinal to transverse cross section $R$, than the proton. This suggests either an unexpected difference in $R$ for the proton and neutron or a suppression of the gluonic distribution in nuclei. 

\end{abstract}


\maketitle

\section{INTRODUCTION}

Nucleon and nuclear structure functions as measured in inclusive
electron scattering have proven to be
 very useful in probing the fundamental, underlying quark dynamics.
Lepton--nucleon scattering experiments at high energy have been conducted over a wide kinematic
range at various experimental facilities such as SLAC, DESY, CERN, Jefferson Lab; see
for instance Refs.\ \cite{dasu,140X,BCDMS,NMC,DESY,ZEUS,vladas}. The data obtained
from these experiments have helped develop the description of hadrons as composite objects of
quarks and gluons which interact weakly at large energy scales but very strongly at low energy scales.
At large enough values of the square of the four--momentum transfer, $Q^2$, corresponding to small
wavelengths of the virtual photon probe, the lepton--nucleon interaction can be viewed as
the incoherent scattering of the virtual photon from a single quark. The experimental results can be interpreted in the framework of perturbative QCD in terms of single parton densities. However, as $Q^2$ decreases, the description of the nucleon's structure becomes more complex, and initial and final state interactions
between the struck quark and the remnants of the target must be included.
The transition from perturbative to non--perturbative QCD is of great interest as it
involves the fundamental dynamics of bound matter, which is not yet well understood.

The kinematics for the inclusive
electron--proton scattering process can be described
 in the one--photon exchange approximation in
terms of the square of the four--momentum transfer ($Q^2=-q^2$, the virtuality of the photon),
and the Bjorken scaling variable
$x=Q^2/{2M\nu}$, where $\nu = E-E'$ is the energy in the target rest frame of the
virtual photon exchanged, $M$ is the proton mass, and $E$ and $E'$ are the incident and scattered electron energies, respectively. In this case the differential cross section for inclusive unpolarized electron scattering can be written as:

\begin{equation}
\label{eq:xsect1}
\frac{1}{\Gamma} \frac{d^2 \sigma}{d \Omega d E'}=\sigma_T + \varepsilon \sigma_L .
\end{equation}

\noindent Here $\sigma_L$ and $\sigma_T$ are the longitudinal and transverse virtual photon absorption
cross sections, respectively, and $\Gamma$ is the transverse virtual photon flux factor,
\begin{equation}
\label{eq:gamma}
\Gamma = \frac{\alpha}{2\pi^2 Q^2}\frac{E}{E'}\frac{\nu (1-x)}{1-\varepsilon},
\end{equation}

\noindent where $\varepsilon$ is the relative longitudinal virtual photon polarization and $\alpha$ is the fine structure function. 
\begin{equation}
\varepsilon = \bigglb[1+2\bigglb(1+\frac{\nu^2}{ Q^2}\biggrb)\tan^2\bigglb(\frac{\theta}{2}\biggrb)\biggrb]^{-1},
\label{eq:epsilon}
\end{equation}

It is convenient to define two dimensionless structure functions, $F_1$ and $F_2$, which are
related to $\sigma_L$ and $\sigma_T$ as follows:
\begin{equation}
\label{eq:f1}
F_1(x, Q^2) = \frac{Q^2}{4\pi^2 \alpha} \frac{(1-x)}{2x}\sigma_T
\end{equation}

\begin{equation}
\label{eq:f2}
F_2(x, Q^2) = \frac{Q^2}{4\pi^2 \alpha}{\frac{1}{1+\frac{Q^2}{\nu^2}}}(1-x)(\sigma_L+\sigma_T).
\end{equation}

While $F_1$ depends only on the transverse virtual absorption cross
section, the more commonly used $F_2$ is a linear combination of
both longitudinal and transverse coupling.
A purely longitudinal structure function can be defined as:

\begin{equation}
\label{eq:fl}
F_L(x, Q^2) = {\frac{Q^2}{4\pi^2 \alpha}}(1-x)\sigma_L.
\end{equation}

This structure function is uniquely sensitive to the gluon
distribution inside the nucleon\ \cite{cooper-sarkar} and can be
written in terms of the gluon density $G(x,Q^2)$ using the
Altarelli--Martinelli equation\ \cite{altarelli78}:


\begin{align}
\label{eq:fl_glue}
F_L(x, Q^2) = {\frac{\alpha}{\pi}}\biggl[{\frac{4}{3}}\int_x^1 {\frac{dy}{y}} \biggl({\frac{x}{y}}\biggr)^2F_2(y, Q^2)\nonumber \\
 + 2\sum\limits_{q}e_q^2 \int_x^1{\frac{dy}{y}} \biggl({\frac{x}{y}}\biggr)^2 \biggl(1- {\frac{x}{y}}\biggr) G(y, Q^2)\biggr]
\end{align}

%

At low $x$ theoretical models predict a dramatic increase in $F_L$ due at least in part
to gluon and quark--antiquark emission. Recent measurements of the longitudinal structure function
from the H1 and ZEUS experiments at HERA\ \cite{DESY,ZEUS} show a non--zero value for $F_L$ at Bjorken $x \leq 0.007$ and large four-momentum transfer squared ($Q^2 \geq 20$~GeV$^2$). These results have sparked renewed theoretical and experimental interest in this structure function (see, for example, Refs.\ \cite{boroun, delgado, golec, ewerz, boroun2013, hentschinski}).

Typically, the ratio of longitudinal and transverse virtual photon absorption cross sections, $R$, is measured to determine $F_L$. $R$ is defined as:

\begin{equation}
\label{eq:r}
R(x, Q^2) = {\frac{\sigma_L}{\sigma_T}} = {\frac{F_L}{2xF_1}},
\end{equation}

\noindent This ratio is expected to vanish at large $Q^2$ and moderate $x$ for scattering from
spin--1/2 pointlike partons, but is nonzero (of the order of 0.1 to 0.3\ \cite{whitlowR})
at low values of $Q^2$ and moderate $x$ due in part to the fact that quarks can carry transverse momentum.

Determining separately the longitudinal and transverse structure
functions, $F_L$ and $F_1$, and thus the ratio $R$, is experimentally challenging. It may be achieved via a Rosenbluth--type separation technique\ \cite{rosenbluth}, using Eq.(\ \ref{eq:xsect1}). This procedure requires high-precision cross-section measurements at the same $x$ and $Q^2$ but different values of $\varepsilon$, which requires in turn measurements at a minimum of two different incident beam energies and scattering angles\ \cite{liang}. Consequently there are far fewer experimental data available for $F_L, F_1$, and $R$ than for the structure function $F_2$.

Measured separated structure functions for the neutron, usually extracted from deuterium
data due to the lack of a free neutron target in nature, are even fewer than those for hydrogen.
Most of the existing deuterium measurements were performed at large four-momentum transfer
and do not show significant differences compared to hydrogen\ \cite{dasu,140X}.
This could be due, however, to $R$ itself being quite small at these kinematics, such that any differences may be buried in the measurement uncertainties. Measurements of $R$ at Jefferson Lab\ \cite{vladas} in the low $Q^2$ regime, where $R$ is larger and differences may appear, show a slight difference between hydrogen and deuterium. This is somewhat unexpected, but not in contradiction with theoretical predictions. There is no requirement in perturbative QCD for $R$ to be the same for the proton and the neutron. While hadron helicity conservation requires that both must go to zero at large $Q^2$, there is no requirement that finite--$Q^2$ corrections/values be identical. However, it is generally assumed that $R_p=R_d$ and that higher twist corrections are identical, e.g. in the extraction of neutron structure function $F_2$ from measurements on the deuteron and proton\ \cite{Arrington09, Arrington12, Hen11, Accardi11}.

\begin{figure}
\begin{center}
\includegraphics[scale=0.4]{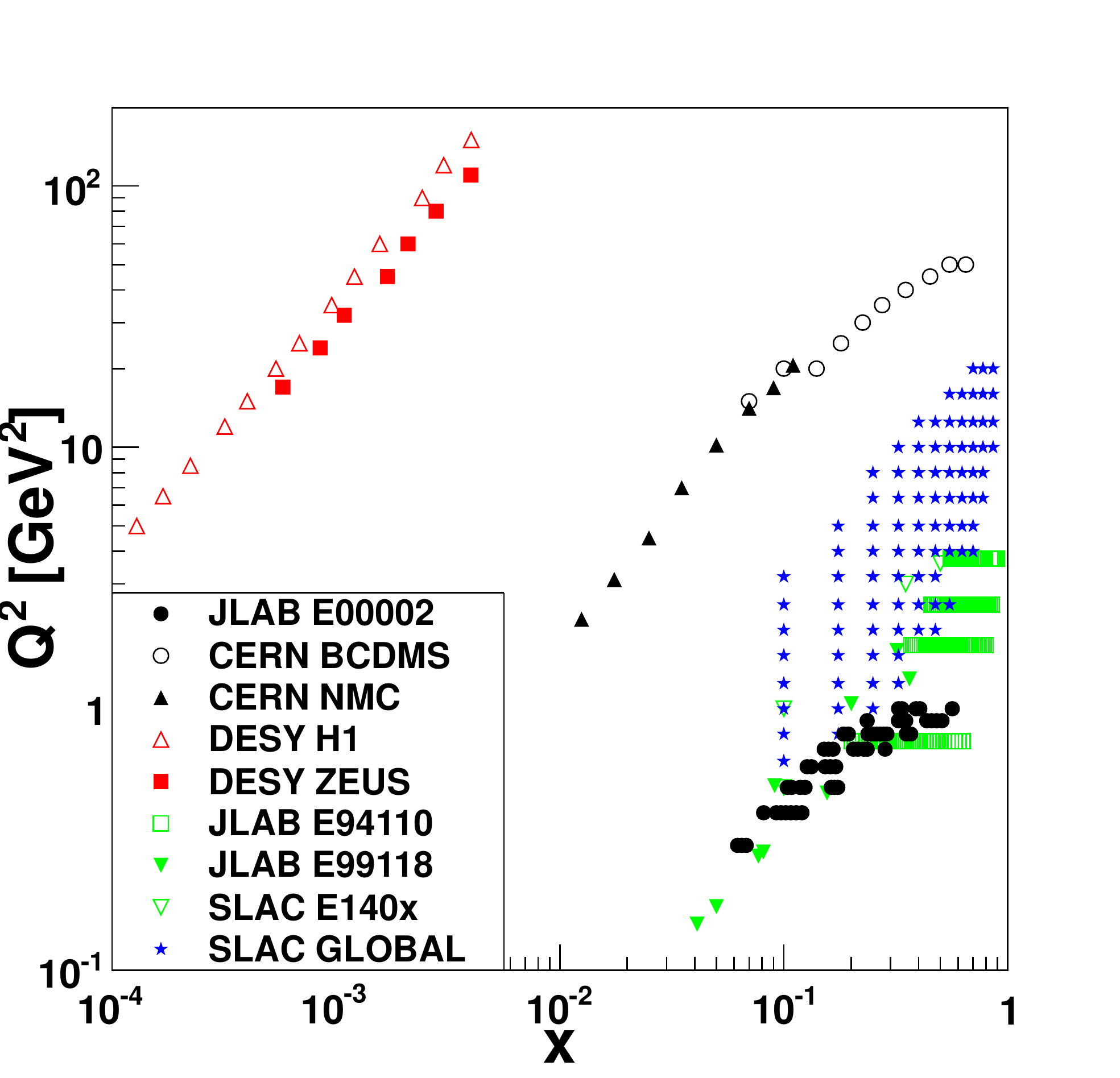}
\end{center}
  \caption{(Color online) Kinematic coverage of world data measurements of the longitudinal structure function for hydrogen: CERN BCDMS\ \protect\cite{BCDMS} open circles, CERN NMC\ \protect\cite{NMC} closed triangles, DESY H1\ \protect\cite{DESY} open triangles, Jefferson Lab E94--110\ \protect\cite{liang}open squares, Jefferson Lab E99--118\ \protect\cite{vladas} inverted closed triangles, SLAC E140X\ \protect\cite{140X} inverted open triangles, SLAC GLOBAL\ \protect\cite{whitlow} stars, and for Jefferson Lab experiment E00--002 (current experiment) closed circles.}
\label{fig:kinem}
\end{figure}

In this paper we present results for $F_L$ and $R$ for the proton and deuteron
at low values of $Q^2$ and intermediate $x$. The kinematic coverage of this experiment
(JLab E00--002) is shown in Fig.\ \ref{fig:kinem} together with the coverage of world data: CERN\ \cite{BCDMS,NMC}, DESY\ \cite{DESY,ZEUS}, Jefferson Lab\ \cite{liang, vladas}, and SLAC\ \cite{whitlow, 140X}. The present data extend the kinematic range in $x$ at fixed low $Q^2$ to allow for more detailed studies of the $x$--dependence, as well as for more deuteron and proton comparisons.

This article is structured in five sections. Sections II and III summarize the experiment and data analysis, Section IV discusses the longitudinal structure function results for hydrogen,
while section V presents the comparison of the hydrogen and deuterium structure functions.
Conclusions are drawn in section VI.

\section{EXPERIMENT}

Experiment E00--002 was carried out at the Thomas Jefferson National Accelerator Facility (Jefferson Lab or JLab) in Hall C. The kinematic range covered was $0.2 < Q^2 < 1.1$~GeV$^2$ and $0.02 < x < 0.6$. Electron--proton and electron--deuteron cross sections were measured using four incident electron beam energies (2.24, 3.04, 4.41, and 5.50~GeV). The experiment
utilized the high luminosity, Continuous Wave (CW) electron beam provided by the Continuous Electron Beam Accelerator Facility (CEBAF). To minimize systematic uncertainties, the beam current 
was kept within a few $\mu A$ of the 40~$\mu A$--nominal value. 
 
The cryogenic targets used were 4--cm long hydrogen (H) and deuterium (D) "tuna can"--shaped aluminum cells. The cryogenic target walls were 0.125~mm thick. The target assembly, described in detail in Refs.\ \cite{simona_paper, eric_elastic}, also included an aluminum "dummy target" that was used for background measurements and subtraction. Precise cross section measurements require accurate knowledge of target thickness and target density. Due to the circular geometry of the cryogenic target cell the effective target length seen by the beam depended upon both the central position of the beam spot and the size and form of the beam raster pattern. The effective target length was determined for Jefferson Lab experiment E00--116 which ran during the same period and details of the analysis are presented in\ \cite{simona_paper}. Density fluctuation studies for the target cells and the beam currents used in this experiment are presented in various publications (see for example Refs.\ \cite{simona_paper, eric_elastic} and amount to $0.35\%/100$~$\mu A$.     

The basic equipment in Hall C at Jefferson Lab consists of two magnetic spectrometers: the High Momentum Spectrometer (HMS) and the Short Orbit Spectrometer (SOS). The HMS was used to detect the scattered electrons, while the SOS was used for detecting positrons which were used to estimate the electron background originating from charge--symmetric processes, such as $\pi^0$ production and subsequent pair symmetric decay. 
The data were taken at HMS angles varying from 11${\deg}$ to 50${\deg}$.  

The HMS is a magnetic spectrometer consisting of a 25${\deg}$ vertical bend dipole magnet for momentum dispersion and three quadrupole magnets for focusing. For this experiment the HMS was operated in point--to--point optical tune. The HMS solid angle is defined mainly by the octagonal collimator and is 6.8~mSr. The HMS momentum acceptance was $\pm 8\%$ while the angular acceptance was $\pm 35$~mrad. The SOS was a resistive $QD\bar D$ magnetic spectrometer with a 9~mSr solid angle. The SOS momentum acceptance was $ -15 \%$ to $+20\%$  while the angular acceptance was $\pm 60$~mrad.   

The HMS and SOS detector packages were very similar consisting of two drift chambers for track
reconstruction (the HMS drift chambers are described in Ref.\ \cite{baker}), scintillator arrays for triggering, and a threshold gas \v Cerenkov and electromagnetic calorimeters\ \cite{hamlet}, which were both used in the present experiment for particle identification and pion rejection. More details on the two spectrometers and their detector packages can be found in Refs.\ \cite{abbott, dutta2003, simona_paper, simona, vladas_thesis}.

\section{DATA ANALYSIS}

The inclusive double differential cross section for each energy and angle bin within the spectrometer acceptance was determined from

\begin{equation}
\label{xsect2}
{\frac{d\sigma}{d\Omega dE'}} = {\frac{Y_{corr}}{L\Delta\Omega \Delta E'}},
\end{equation}

\noindent where $\Delta \Omega (\Delta E')$ is the bin width in solid angle (scattered electron energy), $L$ is the total integrated luminosity, and $Y_{corr} $ is the measured electron yield after accounting for detector inefficiencies, background events mentioned before, and radiative corrections.
%
%

To obtain the Born cross section (the leading order, one photon exchange contribution) the measured yield has to be corrected for the higher order electromagnetic processes which contribute to the inclusive electron--nucleon scattering. These radiative processes can be divided into two main categories: internal, which originate due to the fields of the particles at the scattering vertex (vacuum polarization, vertex corrections, two-photon exchange, and bremsstrahlung emission in the field of the proton from which the scattering took place), and external, which originate due to the fields of particles in the bulk target materials (processes that occur either before or after the primary scattering vertex). The radiative correction factors, which account for these higher order processes, were evaluated using the same procedure used in various Jefferson Lab analyses (see for example Refs.\ \cite{simona_paper, eric_elastic}) and which is based originally on the Mo and Tsai prescription\ \cite{MoTsai1, MoTsai2}. The theoretical uncertainties in the radiative correction procedure
were studied in Ref.\ \cite{Liang_thesis} and were estimated to be 0.5\% point--to--point and 1.0\% normalized.

For every bin in scattering angle and scattered electron energy the cross section
was corrected for the variation of the cross section over the acceptance with
the angle $\theta$ (bin--centering correction) and for radiative effects to yield
the value of the cross section at the central angle. The radiative effects strongly depend on the kinematics. To minimize the dependence on the model\ \cite{eric2010, eric2008} used to compute both the bin--centering correction and the radiative effects, an iterative procedure was
employed. Corrected data were fit to obtain new parameters for the model used, and new corrections were calculated. These steps were repeated until the fit parameters converged.
Statistical and systematic uncertainties for the cross section were at most $\sim$3.5\%.

The total systematic uncertainty in the differential cross section was taken as the quadratic sum of all the systematic uncertainties contributing to the cross section measurement. The total systematic uncertainty not including the one from the radiative corrections varies per kinematic bin between 1.19\% and 1.43\% (see Table~I).

The extraction of the longitudinal structure function was accomplished via
the Rosenbluth technique, where measurements are made for two or more values of $\varepsilon$ at fixed Bjorken $x$ and $Q^2$, and the reduced cross section is fit linearly as a function of $\varepsilon$, as in Eq.(\ref{eq:xsect1}). Since even after all corrections are applied the data at various $\varepsilon$ are not at the exact same $x$ and $Q^2$ as needed for the Rosenbluth separation method, the data were interpolated toward a common $x$ and $Q^2$ using the cross section parameterization\ \cite{eric2010,eric2008}.

In a Rosenbluth separation one needs to distinguish between uncertainties that are correlated between measurements at different $\varepsilon$, such as uncertainties in target thickness and integrated charge, and uncorrelated ones such as spectrometer acceptance or background subtraction at different angles. Not including the contributions from radiative corrections, the uncorrelated systematic uncertainties in the cross-section measurements in this experiment amounted to $\sim$0.88\%.

\begin{table}[h!]
\label{tab:sys}
\begin{center}
\begin{tabular}{c c }
\hline
Quantity & Uncertainty \\ \hline \hline
 Beam Energy & 0.30\%  \\
Scattered electron Energy& 0.25\%\\
Scattered electron Angle& 0.30\% \\
Target Density& 0.35\%\\
Dead Time Corrections& 0.25\% \\
Tracking Efficiency& 0.25\%\\
 Efficiency (\u{C}er. Cal.)& 0.35\% \\
 Charge Symmetric Background & 0.20\% \\
 Beam Charge& 0.71--1.07\% \\
 Acceptance& 0.50\%\\\hline
 TOTAL     & 1.19--1.43\% \\\hline \hline
\end{tabular}
\end{center}
\caption{Systematic uncertainties in the differential cross section.}
\end{table}



Both $\varepsilon$ and $\Gamma$ were calculated from the
measured kinematic variables using Eqs.(\ref{eq:epsilon}) and (\ref{eq:gamma}), respectively.
 The intercept of the fit is the transverse cross section $\sigma_T$ (and therefore the structure function $F_1(x,Q^2)$), while the slope is the longitudinal cross section $\sigma_L$, from which
the structure function $F_L(x,Q^2)$ can be calculated by using Eq.(\ref{eq:fl}).


The radiative corrections do not include the contribution of hard 
two-photon exchange (TPE). While this appears to have a significant impact on 
Rosenbluth extractions of $G_E^p/G_M^p$ for electron--proton elastic scattering at high $Q^2$\ \cite{Arrington07, Puckett11, Puckett12,Tvaskis06}, calculations and recent data suggest that they are typically 
below 2\% for elastic scattering at the $Q^2$ values of this measurement\ \cite{Arrington07, Carlson07,  Arrington11, Rachek14, Adikaram14}. Estimates of the TPE contribution to resonance production\ \cite{Pascalutsa06, Kondratyuk06} 
suggest that the corrections for the resonance region are  smaller than for elastic scattering, e.g. by roughly a factor of two for Delta production. 

For the present data, the longitudinal cross section is typically 20-50\% of the total cross section, so a change in the slope of $\leq$1\% would translate into a 2-5\% correction to $\sigma_L$. Even though a definitive calculation of the two--photon effect is not available, this correction is typically smaller than the total statistical and systematic uncertainties, and thus should have minimial impact on the final results. $F_L$ is only sensitive to non--linearities caused by the effect. Moreover, there may be some cancellation between the impact on $R$ when comparing $R_D$ and $R_H$.

\section{PROTON LONGITUDINAL STRUCTURE FUNCTION}

\begin{figure}[htp]
\begin{center}
\includegraphics[scale=0.43]{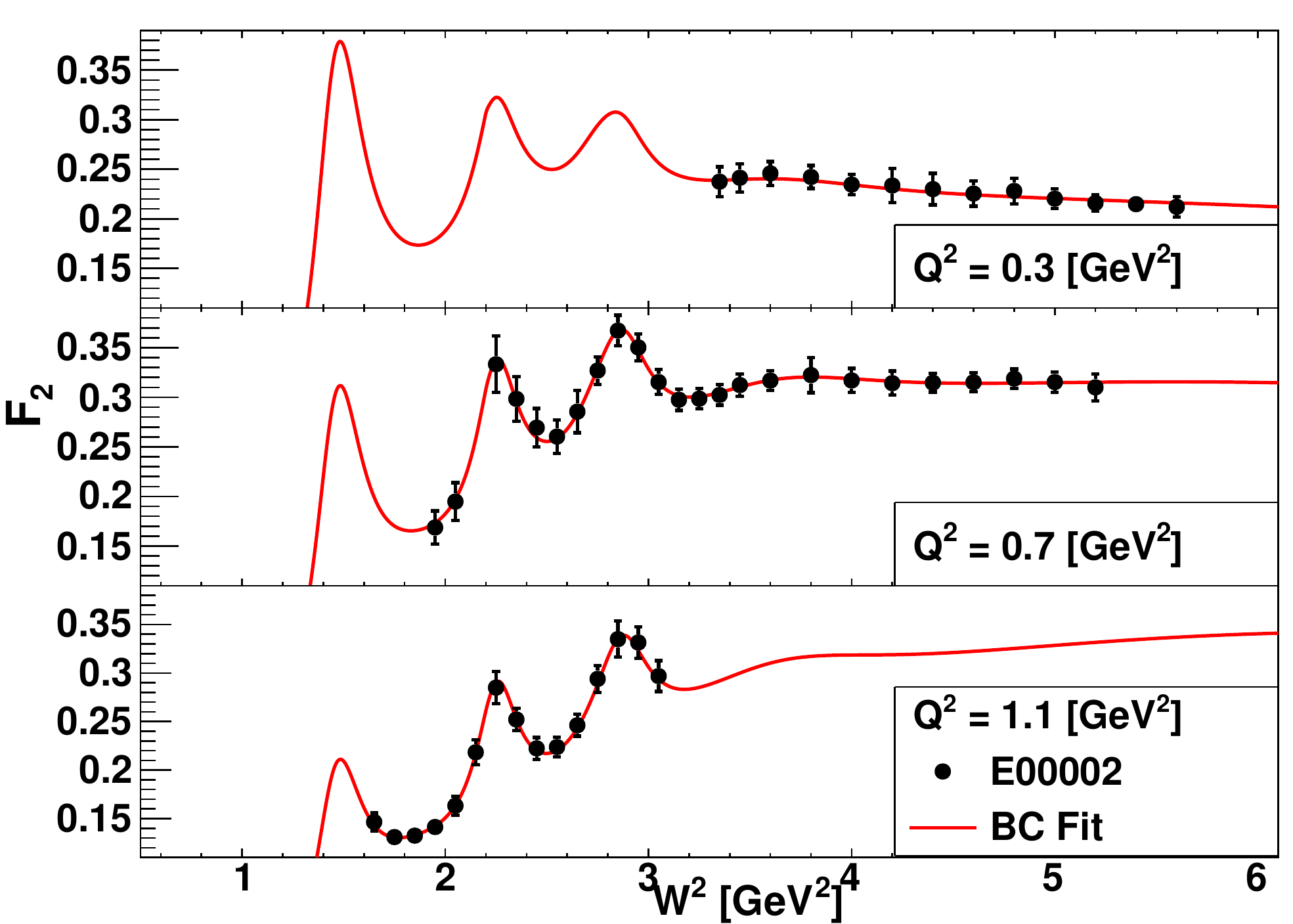}
\end{center}
  \caption{(Color online) $F_2$ structure function data for the proton as a function of invariant mass $W^2$. The three panels correspond to three different $Q^2$ regions: $Q^2 = 0.3$~GeV$^2$ (top panel), $Q^2 = 0.7$~GeV$^2$ (middle panel), $Q^2 = 1.1$~GeV$^2$ (bottom panel). The data are compared with the Bosted--Christy (BC)  parameterization\ \protect\cite{eric2010} (solid curve).}
\label{fig:f2_vs_w2}
\end{figure}

The data from experiment E00--002 were used to extract the proton $F_2$ structure function as well as the proton longitudinal structure function, $F_L$. Three representative spectra depicting the $F_2$ structure function as a function of the invariant mass, $W^2$, are shown in Fig.\ \ref{fig:f2_vs_w2}. The data are compared with the Bosted--Christy (BC) parameterization\ \cite{eric2010}, an empirical fit of inelastic electron--proton cross sections in the resonance region. The agreement between the data and this global fit is very good. 

\begin{figure}[htp]
\begin{center}
\includegraphics[scale=0.43]{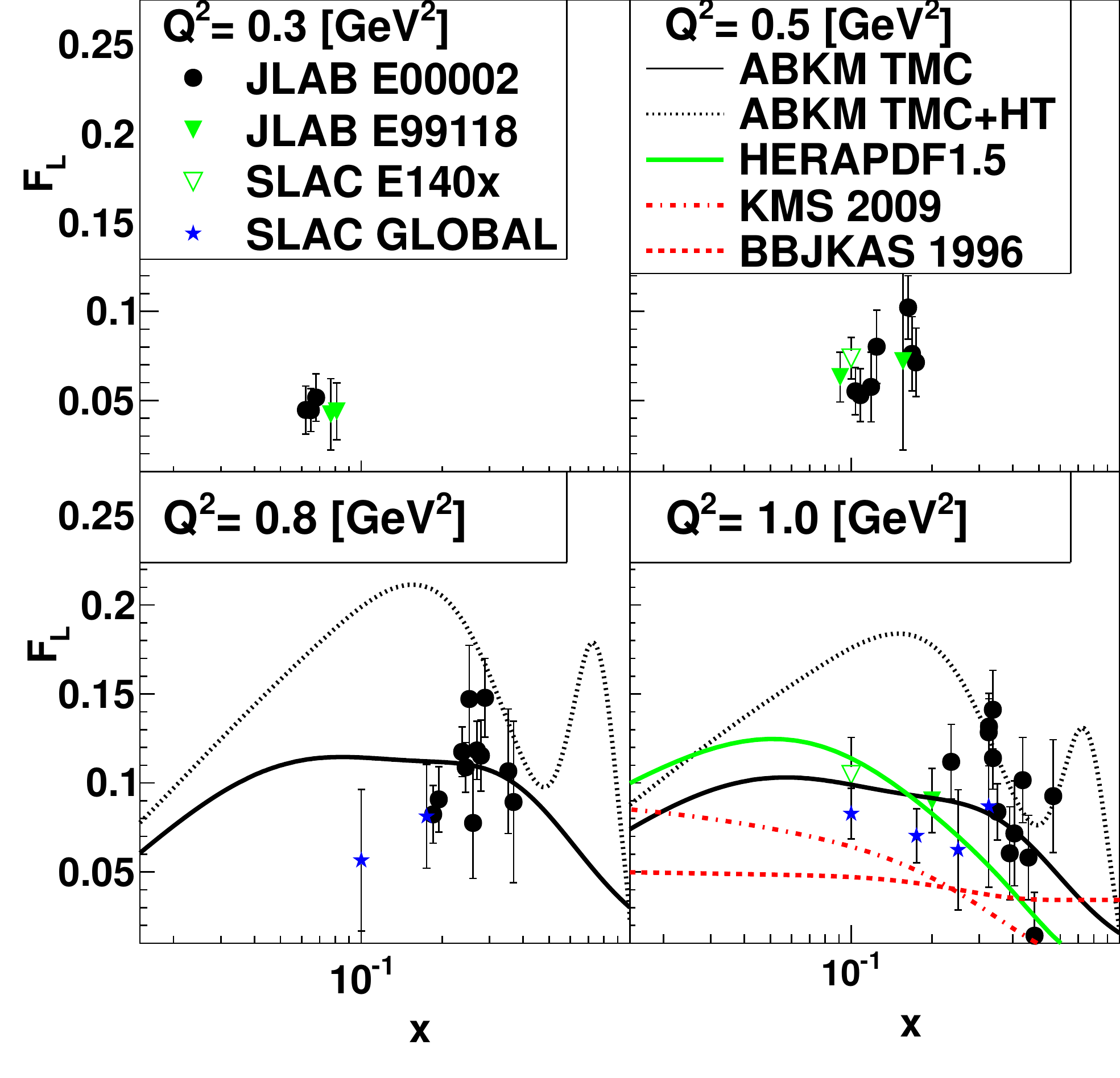}
\end{center}
  \caption{(Color online) Longitudinal structure function data for the proton for fixed $Q^2$ as a function of Bjorken $x$. The four panels correspond to four different $Q^2$ regions: $Q^2 = 0.3$~GeV$^2$ (upper left), $Q^2 = 0.5$~GeV$^2$ (upper right), $Q^2 = 0.8$~GeV$^2$ (lower left), and $Q^2 = 1.0$~GeV$^2$ (lower right). In this kinematic regime data are from Jefferson Lab\ \protect\cite{vladas,liang} and SLAC\ \protect\cite{whitlow,140X}. These are shown with the following parameterizations for comparison: ABKM\ \protect\cite{ABKM,abm11} solid and dotted black curves, HERAPDF1.5\ \protect\cite{HERAPDF} (solid green line), KMS\ \protect\cite{golec} dashed-dotted red curves, and BBJKAS\ \protect\cite{badelek} dashed red curve.}
\label{fig:fig1}
\end{figure}

The results for the proton $F_2$ structure function, the longitudinal structure function $F_L$ and the ratio $R$ are presented in Table II.   
The proton longitudinal structure function $F_L$ is shown in Fig.\ \ref{fig:fig1}
as a function of Bjorken $x$ (for $x >$ 0.01) for four values of $Q^2$ = 0.30, 0.50, 0.80 and 1.0~GeV$^2$, respectively. Data are included from this experiment (E00--002, closed black circles), the JLab experiment E99--118\ \cite{vladas} (inverted closed triangles), SLAC experiment E140X\ \cite{140X} (inverted open triangles), and a global analysis of SLAC data\ \cite{whitlow} (closed stars).

\begin{table*}[h!]
\label{tab:data_proton}
\begin{center}
\begin{tabular}{ c c c c c c c c c c c }
\hline\hline
\hskip 0.15 in $Q^2$ \hskip 0.15 in & \hskip 0.15 in $x$ \hskip 0.15 in & \hskip 0.15 in $R$ \hskip 0.15 in & \hskip 0.15 in$\Delta_{stat}$ \hskip 0.15 in &\hskip 0.15 in $\Delta_{total}$ \hskip 0.15 in& \hskip 0.15 in$F_2$\hskip 0.15 in & \hskip 0.15 in$\Delta_{stat}$\hskip 0.15 in & \hskip 0.15 in$\Delta_{total}$\hskip 0.15 in & \hskip 0.15 in$F_L$ \hskip 0.15 in&\hskip 0.15 in $\Delta_{stat}$ \hskip 0.15 in&\hskip 0.15 in $\Delta_{total}$ \hskip 0.15 in\\ \hline\hline
     &       &        &        &        &        &        &        &        &        &   \\
0.30 & 0.062 & 0.2437 & 0.0757 & 0.0843 & 0.2178 & 0.0135 & 0.0148 & 0.0446 & 0.0123 & 0.0135\\
0.30 & 0.065 & 0.2418 & 0.0681 & 0.0739 & 0.2178 & 0.0120 & 0.0135 & 0.0445 & 0.0113 & 0.0121\\
0.30 & 0.068 & 0.2803 & 0.0788 & 0.0831 & 0.2235 & 0.0137 & 0.0150 & 0.0516 & 0.0126 & 0.0133\\
0.40 & 0.081 & 0.2037 & 0.0656 & 0.0724 & 0.2481 & 0.0139 & 0.0156 & 0.0444 & 0.0130 & 0.0141\\ 
0.40 & 0.093 & 0.3420 & 0.0804 & 0.0823 & 0.2650 & 0.0153 & 0.0171 & 0.0727 & 0.0144 & 0.0148\\
0.40 & 0.097 & 0.3005 & 0.0992 & 0.1003 & 0.2621 & 0.0201 & 0.0215 & 0.0656 & 0.0181 & 0.0183\\
0.40 & 0.102 & 0.3154 & 0.0915 & 0.0926 & 0.2657 & 0.0183 & 0.0198 & 0.0696 & 0.0168 & 0.0171\\
0.40 & 0.108 & 0.3665 & 0.1258 & 0.1266 & 0.2695 & 0.0242 & 0.0254 & 0.0797 & 0.0217 & 0.0219\\
0.40 & 0.114 & 0.3302 & 0.1313 & 0.1319 & 0.2692 & 0.0266 & 0.0277 & 0.0744 & 0.0238 & 0.0240\\
0.40 & 0.121 & 0.3441 & 0.0672 & 0.0697 & 0.2778 & 0.0137 & 0.0158 & 0.0802 & 0.0129 & 0.0135\\
0.50 & 0.104 & 0.2271 & 0.0564 & 0.0608 & 0.2768 & 0.0128 & 0.0151 & 0.0551 & 0.0125 & 0.0133\\
0.50 & 0.108 & 0.2149 & 0.0650 & 0.0674 & 0.2766 & 0.0152 & 0.0171 & 0.0530 & 0.0145 & 0.0149\\
0.50 & 0.119 & 0.2283 & 0.0868 & 0.0883 & 0.2819 & 0.0205 & 0.0221 & 0.0576 & 0.0193 & 0.0196\\
0.50 & 0.124 & 0.3303 & 0.1010 & 0.1020 & 0.2912 & 0.0217 & 0.0232 & 0.0802 & 0.0204 & 0.0206\\
0.50 & 0.163 & 0.4154 & 0.0909 & 0.0919 & 0.2933 & 0.0180 & 0.0198 & 0.1022 & 0.0175 & 0.0179\\
0.50 & 0.168 & 0.2891 & 0.0926 & 0.0932 & 0.2830 & 0.0205 & 0.0221 & 0.0761 & 0.0207 & 0.0209\\
0.50 & 0.174 & 0.2616 & 0.0828 & 0.0834 & 0.2835 & 0.0194 & 0.0210 & 0.0714 & 0.0191 & 0.0193\\
0.60 & 0.127 & 0.3225 & 0.0551 & 0.0575 & 0.3042 & 0.0126 & 0.0153 & 0.0812 & 0.0117 & 0.0122\\
0.60 & 0.133 & 0.3206 & 0.0569 & 0.0589 & 0.3095 & 0.0132 & 0.0158 & 0.0829 & 0.0124 & 0.0129\\
0.60 & 0.153 & 0.3329 & 0.1076 & 0.1083 & 0.3070 & 0.0244 & 0.0259 & 0.0872 & 0.0232 & 0.0234\\
0.60 & 0.161 & 0.4497 & 0.1303 & 0.1309 & 0.3129 & 0.0261 & 0.0276 & 0.1119 & 0.0249 & 0.0251\\
0.60 & 0.171 & 0.5026 & 0.1446 & 0.1452 & 0.3209 & 0.0279 & 0.0293 & 0.1256 & 0.0270 & 0.0272\\
0.70 & 0.152 & 0.3214 & 0.0652 & 0.0671 & 0.3185 & 0.0154 & 0.0178 & 0.0864 & 0.0150 & 0.0155\\
0.70 & 0.158 & 0.2303 & 0.0495 & 0.0511 & 0.3142 & 0.0133 & 0.0160 & 0.0662 & 0.0124 & 0.0128\\
0.70 & 0.166 & 0.2741 & 0.0550 & 0.0560 & 0.3131 & 0.0139 & 0.0165 & 0.0767 & 0.0131 & 0.0134\\
0.70 & 0.205 & 0.3283 & 0.0715 & 0.0731 & 0.3120 & 0.0163 & 0.0186 & 0.0934 & 0.0174 & 0.0179\\
0.70 & 0.214 & 0.5300 & 0.2343 & 0.2346 & 0.3156 & 0.0431 & 0.0440 & 0.1345 & 0.0433 & 0.0435\\
0.70 & 0.228 & 0.3642 & 0.0842 & 0.0849 & 0.2998 & 0.0181 & 0.0200 & 0.1010 & 0.0188 & 0.0191\\
0.70 & 0.236 & 0.3830 & 0.0839 & 0.0846 & 0.3056 & 0.0180 & 0.0199 & 0.1083 & 0.0190 & 0.0192\\
0.70 & 0.283 & 0.2666 & 0.1282 & 0.1285 & 0.2878 & 0.0303 & 0.0314 & 0.0851 & 0.0343 & 0.0344\\
0.80 & 0.185 & 0.2863 & 0.0643 & 0.0652 & 0.3215 & 0.0161 & 0.0185 & 0.0824 & 0.0159 & 0.0162\\
0.80 & 0.194 & 0.3224 & 0.0772 & 0.0780 & 0.3194 & 0.0184 & 0.0205 & 0.0908 & 0.0182 & 0.0184\\
0.80 & 0.237 & 0.4243 & 0.0617 & 0.0630 & 0.3160 & 0.0131 & 0.0158 & 0.1175 & 0.0135 & 0.0139\\
0.80 & 0.245 & 0.3960 & 0.0623 & 0.0633 & 0.3037 & 0.0130 & 0.0156 & 0.1088 & 0.0136 & 0.0140\\
0.80 & 0.252 & 0.6037 & 0.1796 & 0.1801 & 0.3054 & 0.0295 & 0.0308 & 0.1472 & 0.0299 & 0.0301\\
0.80 & 0.261 & 0.2579 & 0.1243 & 0.1245 & 0.2912 & 0.0305 & 0.0316 & 0.0776 & 0.0311 & 0.0312\\
0.80 & 0.269 & 0.3911 & 0.0680 & 0.0687 & 0.3188 & 0.0151 & 0.0176 & 0.1183 & 0.0160 & 0.0164\\
0.80 & 0.279 & 0.3273 & 0.0688 & 0.0693 & 0.3486 & 0.0181 & 0.0206 & 0.1154 & 0.0196 & 0.0200\\
0.80 & 0.289 & 0.4226 & 0.0811 & 0.0817 & 0.3642 & 0.0199 & 0.0224 & 0.1479 & 0.0217 & 0.0221\\
0.80 & 0.353 & 0.3056 & 0.1238 & 0.1240 & 0.2943 & 0.0287 & 0.0299 & 0.1066 & 0.0349 & 0.0350\\
0.80 & 0.369 & 0.2103 & 0.1236 & 0.1237 & 0.3216 & 0.0358 & 0.0369 & 0.0893 & 0.0453 & 0.0454\\
0.90 & 0.236 & 0.3871 & 0.0896 & 0.0902 & 0.3293 & 0.0204 & 0.0224 & 0.1119 & 0.0207 & 0.0210\\
0.90 & 0.325 & 0.4089 & 0.0764 & 0.0770 & 0.3133 & 0.0163 & 0.0185 & 0.1285 & 0.0185 & 0.0189\\
0.90 & 0.337 & 0.4191 & 0.0830 & 0.0835 & 0.2672 & 0.0149 & 0.0167 & 0.1140 & 0.0172 & 0.0175\\
0.90 & 0.350 & 0.3033 & 0.0700 & 0.0704 & 0.2433 & 0.0135 & 0.0151 & 0.0838 & 0.0157 & 0.0158\\
0.90 & 0.435 & 0.4568 & 0.1472 & 0.1475 & 0.1862 & 0.0178 & 0.0185 & 0.1016 & 0.0238 & 0.0240\\
0.90 & 0.457 & 0.2564 & 0.1262 & 0.1263 & 0.1568 & 0.0169 & 0.0175 & 0.0581 & 0.0236 & 0.0237\\
0.90 & 0.481 & 0.05490 & 0.0986 & 0.0987 & 0.1431 & 0.0168 & 0.0173 & 0.0142 & 0.0244 & 0.0244\\
0.90 & 0.509 & 0.03044 & 0.1183 & 0.1183 & 0.1460 & 0.0212 & 0.0216 & 0.0087 & 0.0329 & 0.0329\\
1.00 & 0.326 & 0.3912 & 0.0696 & 0.07019 & 0.3408 & 0.0166 & 0.0191 & 0.1317 & 0.0184 & 0.0188\\
1.00 & 0.337 & 0.4124 & 0.0815 & 0.0820 & 0.3456 & 0.0191 & 0.0215 & 0.1412 & 0.0217 & 0.0221\\
1.00 & 0.389 & 0.2074 & 0.1025 & 0.1027 & 0.2297 & 0.0209 & 0.0219 & 0.0605 & 0.0261 & 0.0262\\
1.00 & 0.405 & 0.2063 & 0.0969 & 0.0970 & 0.2656 & 0.0229 & 0.0241 & 0.0717 & 0.0294 & 0.0295\\
1.00 & 0.565 & 0.3706 & 0.1663 & 0.1665 & 0.1612 & 0.0196 & 0.0201 & 0.0926 & 0.0316 & 0.0317 \\ \hline\hline
\end{tabular}
\end{center}
\caption{Proton structure functions $F_2$, $R$, and $F_L$ extracted using the Rosebluth separation technique.}
\end{table*}


At these low values of $Q^2$ there are no other separated data within this range of $x$, as can be seen from Fig.\ \ref{fig:kinem}. The results for the deuteron differ from those for the proton and will be discussed in section IV.

The curves in Fig.\ \ref{fig:fig1} represent the ABKM\ \cite{ABKM,abm11} (solid and dotted black lines, adding respectively target mass
and higher twist), the HERAPDF1.5 NNLO QCD global parton distribution (PDF) fit (solid green line)\ \cite{HERAPDF}, and two $k_T$ factorization (dasehd--dotted and dashed red lines) parameterizations\ \cite{golec, badelek}. The ABKM parameterization is a next--to--next--to--leading--order (NNLO) PDF fit obtained by a global analysis of available hard--scattering data in the ${\overline {MS}}$ factorization scheme. 
The ABKM parameterization of the structure function includes
terms that take into account non-perturbative target-mass corrections (TMC) and
higher-twist effects\ \cite{abm11}. Therefore, the curves in Fig.\ \ref{fig:fig1} are
shown including TMC, and both without (solid black line)
and with (dotted black line) the inclusion of higher--twist effects. 

These curves are only shown for the two
higher--$Q^2$ panels because of constraints on the region of validity for the PDF fits.
For the two higher--$Q^2$ panels ($Q^2=0.8$ and $Q^2=1.0$~GeV$^2$), where the parameterizations become better constrained by existing structure-function data, there is reasonable agreement
between the models and the data. The global fits typically employ $F_2$ and not $F_L$ data, and extract gluon information not from $F_L$ via Eq.(\ \ref{eq:fl_glue}) but rather from the $Q^2$ evolution of $F_2$. 
Hence, even if deviations between the data and the curves are noticeable, the general agreement at the low $Q^2$ of the $F_L$ data is unexpected, given that the parameterizations originate
from perturbative methods. For the kinematic regime of the Jefferson Lab data the target-mass corrections and higher-twist effects are expected to be significant and thus improved agreement is expected with
parameterizations that include these effects. However, the ABKM inclusion of such effects seems to be unnecessary and yields significantly worse agreement with the data.
It is important to note also that $F_L$ is dominated by the gluon $g(x)$ PDF which has
large uncertainties ($\sim 20$\% or more in this $x$ range) that are not shown here.

The PDF fits discussed above utilize the standard collinear
factorization formalism. Another approach in modeling $F_L$ is to employ the
$k_T$ factorization theorem, which corresponds to the (virtual) photon--gluon
fusion mechanism. Here, the gluon is off--shell with its virtuality dominated
by the transverse momentum $k_T$\ \cite{golec}.
This model (KMS) is shown as a dashed--dotted line at $Q^2 = 1$~GeV/$^2$
in Fig.\ \ref{fig:fig1}. While it undercuts the data, it should be noted that it
was developed specifically to describe the low $x$, higher $Q^2$ regime of
data obtained at DESY. An older version of this photon--gluon fusion model\ \cite{badelek1996,badelek}, which specifically included higher twist effects,
is also shown in Fig.\ \ref{fig:fig1} (dashed line, BBJKAS), which shows increased strength at
larger $x$. A new $k_T$ factorization model that also includes higher twist is in
development\ \cite{stasto_private}.

\begin{figure}[t]
\begin{center}
\includegraphics[scale=0.43]{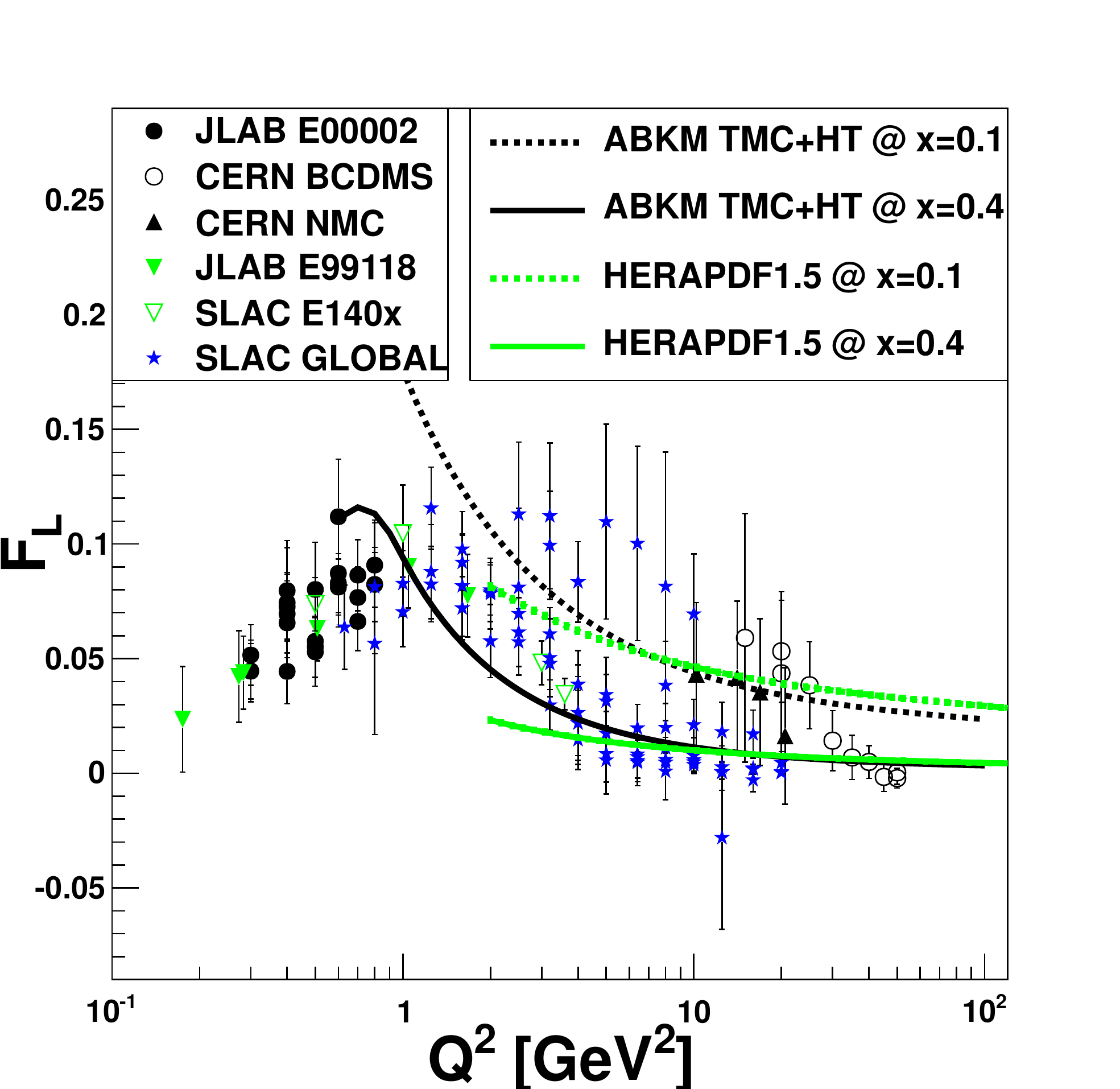}
\end{center}
  \caption{ (Color online) Longitudinal structure-function data for the proton as a function of $Q^2$
for $x\geq 0.05$. Data are from Refs.\ \protect\cite{NMC, BCDMS, 140X, vladas, liang, whitlow}.
The solid and dashed lines represent parameterizations discussed
in the text. The dotted lines correspond to $x=0.1$, while the solid line corresponds to $x=0.4$.}
\label{fig:fig2}
\end{figure}

\begin{figure}[t]
\begin{center}
\includegraphics[scale=0.43]{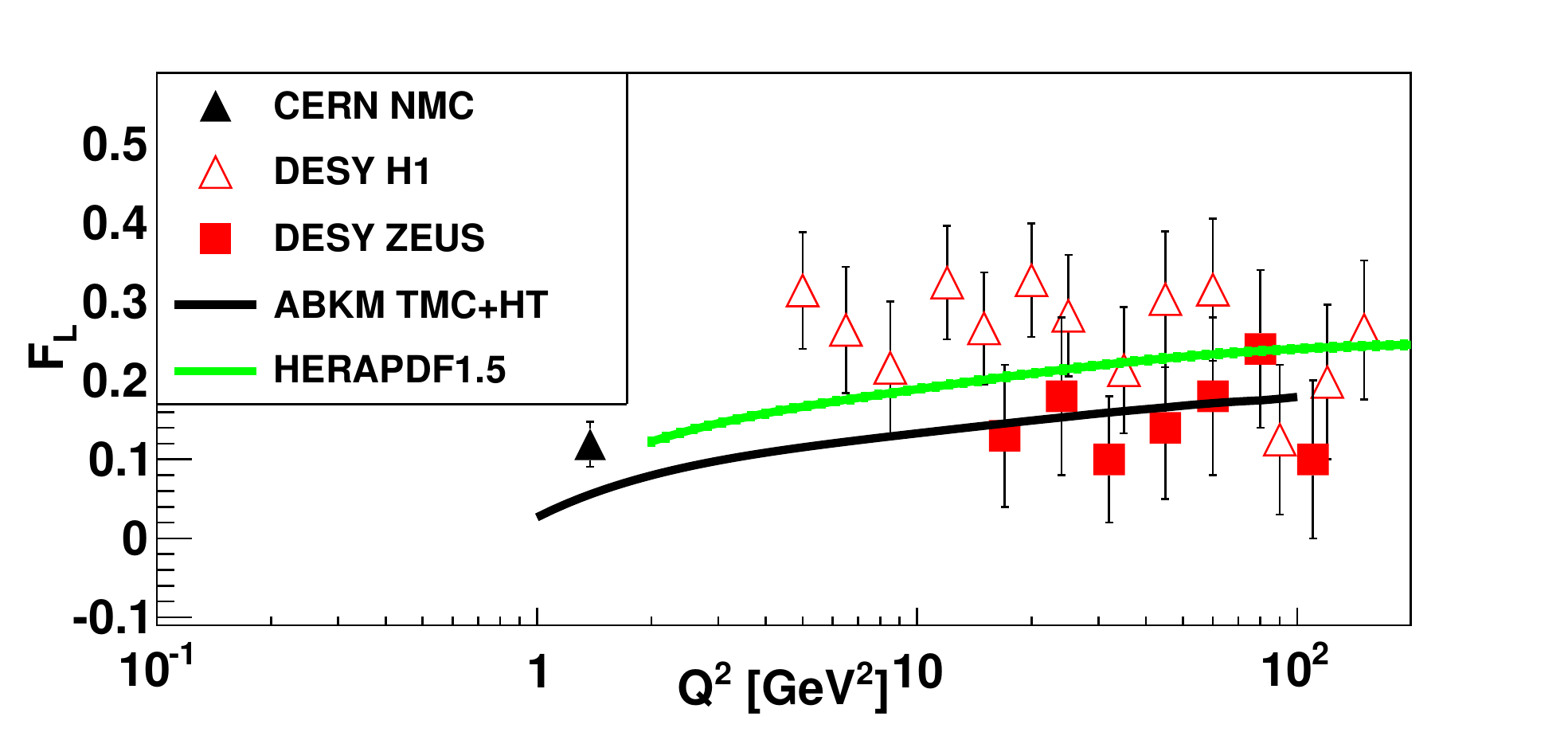}
\end{center}
  \caption{ (Color online) Proton longitudinal structure--function data at very low x: $0.0001 \leq x \leq 0.006$ as a function of $Q^2$. Data are from Refs.\ \protect\cite{DESY, NMC}. The solid lines represent parameterizations discussed in the text and evaluated at $x=0.004$.}
\label{fig:desy_fl}
\end{figure}

To further investigate the $Q^2$ behavior of the data, we present
in Fig.\ \ref{fig:fig2} and\ \ref{fig:desy_fl} the world data for $F_L$ versus $Q^2$ for $x$ above 0.05 (Fig.\ \ref{fig:fig2}) and for $x$ between 0.0001 and 0.006 (Fig.\ \ref{fig:desy_fl}).
The data in Fig.\ \ref{fig:fig2} were obtained from various experiments at CERN (NMC and BCDMS) and from 
lower--energy SLAC and JLab experiments (E140x, E99--118, and E00--002). The data in Fig.\ \ref{fig:desy_fl} are primarily from the recent DESY H1 and ZEUS experiments\ \cite{DESY,ZEUS}. In this latter kinematic regime the longitudinal structure function $F_L$ is dominated by the gluon parton distribution function and is well described by the PDF parameterizations which do include these data.
The range of the CERN, SLAC, and JLab data, Fig.\ \ref{fig:fig2}, includes regions where meson--cloud or sea-quark effects may become relevant, as well as the larger $x$ valence region where
non--perturbative binding and other effects may also become relevant, particularly at lower $Q^2$. The ABKM and HERAPDF fits described previously are also shown in Fig.\ \ref{fig:fig2} at $x=0.1$ and $x=0.4$, and in Fig.\ \ref{fig:desy_fl} for $x=0.0004$.

The change which can be seen in Fig.\ \ref{fig:fig2} in the $Q^2$-dependence of the higher $x$  $F_L$ structure--function data from low ($<$1) to high $Q^2$ ($>$1~GeV$^2$) is of particular interest. Due to current conservation the interaction of longitudinal virtual photons vanishes in the real photon $Q^2=0$ limit. Hence, since $\sigma_L \approx F_L/Q^2$, one expects $F_L$ to vanish as $Q^4$ when $Q^2 \rightarrow 0$.
This behavior has been used as a kinematic constraint for theoretical modeling\ \cite{golec,stasto09}.
The moderate to high $x$ data depicted here were obtained far from the strict real photon $Q^2$=0 limit,
however, and yet nonetheless seem to converge smoothly below $Q^2\approx 1$~GeV$^2$ for all $x$.

In comparison, in the range $1< Q^2< 10$~GeV$^2$, there is a significant spread to the data. This spread is due largely to the expected $x$--dependence as these data span the range
$0.1\leq x \leq 0.7$. This spread is predicted also in the global fits at differing values of $x$.  As discussed previously, the non--perturbative effects in the transition to lower $Q^2$ are overestimated in the ABKM analysis, which regardless agrees reasonably well with the data if the average $x$ is assumed to be $0.4$. The average $x$ for the data in the range $0.5< Q^2< 5$~GeV$^2$ is about 0.2. At the highest values of $Q^2$, above $10$~GeV$^2$,  the data once again converge towards $0$. Here, the structure function $F_L$ must vanish due to hadron helicity conservation.

\section{ PROTON/DEUTERON COMPARISON}

Lastly, we compare the data accumulated for the deuteron to that for the proton within the $Q^2$ range accessible to the SLAC and JLab experiments. Inclusive electron
scattering experiments such as E00--002 are able to alternately acquire proton and
deuteron data, with reduced point--to--point systematic uncertainties in the
ratio $R$ = $\sigma_L/\sigma_T$, as some sources of uncertainty are common and therefore
vanish in the ratio. It is traditional that Rosenbluth--type measurements present the
quantity $R$ = $\sigma_L/\sigma_T$, in part for systematic uncertainty evaluation and in
part to extract cross sections via a variant of Eq.\ \ref{eq:xsect1}:
\begin{equation}
\label{eq:xsect2}
\frac{1}{\Gamma} \frac{d^2 \sigma}{d \Omega d E'} = \sigma_T(1+ \varepsilon R).
\end{equation}

\noindent Since $\varepsilon$ and $R$ are typically small, $\sigma_T$ and hence $F_1$ account
for 70--90\% of both the cross section and the $F_2$ structure function.
%
%

The results for the proton $F_2$ structure function, the longitudinal structure function $F_L$ and the ratio $R$ are presented in Table III.
\begin{table*}[h!]
\label{tab:data_deut}
\begin{center}
\begin{tabular}{ c c c c c c c c c c c }
\hline\hline
\hskip 0.15 in $Q^2$ \hskip 0.15 in & \hskip 0.15 in $x$ \hskip 0.15 in & \hskip 0.15 in $R$ \hskip 0.15 in & \hskip 0.15 in$\Delta_{stat}$ \hskip 0.15 in &\hskip 0.15 in $\Delta_{total}$ \hskip 0.15 in& \hskip 0.15 in$F_2$\hskip 0.15 in & \hskip 0.15 in$\Delta_{stat}$\hskip 0.15 in & \hskip 0.15 in$\Delta_{total}$\hskip 0.15 in & \hskip 0.15 in$F_L$ \hskip 0.15 in&\hskip 0.15 in $\Delta_{stat}$ \hskip 0.15 in&\hskip 0.15 in $\Delta_{total}$ \hskip 0.15 in\\ \hline\hline
     &       &        &        &        &        &        &        &        &        &   \\
0.30 & 0.062 & 0.2103 & 0.0669 & 0.0694 & 0.2030 & 0.0115 & 0.0129 & 0.0369 & 0.0106 & 0.0110\\
0.30 & 0.065 & 0.2141 & 0.0593 & 0.0612 & 0.2039 & 0.0101 & 0.0117 & 0.0377 & 0.0095 & 0.0098\\
0.30 & 0.068 & 0.2142 & 0.0788 & 0.0799 & 0.2061 & 0.0140 & 0.0151 & 0.0383 & 0.0125 & 0.0127\\
0.40 & 0.081 & 0.2367 & 0.0639 & 0.0664 & 0.2384 & 0.0124 & 0.0142 & 0.0483 & 0.0118 & 0.0122\\ 
0.40 & 0.093 & 0.2890 & 0.0711 & 0.0720 & 0.2420 & 0.0133 & 0.0150 & 0.0584 & 0.0124 & 0.0126\\
0.40 & 0.097 & 0.3172 & 0.1017 & 0.1023 & 0.2424 & 0.0187 & 0.0199 & 0.0632 & 0.0168 & 0.0169\\
0.40 & 0.102 & 0.3615 & 0.0976 & 0.0982 & 0.2451 & 0.0170 & 0.0183 & 0.0710 & 0.0156 & 0.0158\\
0.40 & 0.108 & 0.3062 & 0.1169 & 0.1173 & 0.2448 & 0.0221 & 0.0232 & 0.0632 & 0.0199 & 0.0200\\
0.40 & 0.114 & 0.3062 & 0.1263 & 0.1266 & 0.2468 & 0.0242 & 0.0252 & 0.0644 & 0.0217 & 0.0218\\
0.40 & 0.121 & 0.3448 & 0.0584 & 0.0599 & 0.2519 & 0.0109 & 0.0130 & 0.0728 & 0.0100 & 0.0104\\
0.50 & 0.104 & 0.2608 & 0.0552 & 0.0571 & 0.2642 & 0.0115 & 0.0137 & 0.0588 & 0.0112 & 0.0116\\
0.50 & 0.108 & 0.2053 & 0.0601 & 0.0613 & 0.2575 & 0.0132 & 0.0152 & 0.0475 & 0.0126 & 0.0128\\
0.50 & 0.119 & 0.2528 & 0.0833 & 0.0840 & 0.2609 & 0.0177 & 0.0192 & 0.0579 & 0.0166 & 0.0167\\
0.50 & 0.124 & 0.2869 & 0.0903 & 0.0909 & 0.2651 & 0.0187 & 0.0202 & 0.0655 & 0.0175 & 0.0176\\
0.50 & 0.163 & 0.3567 & 0.0726 & 0.0733 & 0.2621 & 0.0139 & 0.0158 & 0.0818 & 0.0132 & 0.0135\\
0.50 & 0.168 & 0.3013 & 0.0738 & 0.0744 & 0.2552 & 0.0148 & 0.0165 & 0.0709 & 0.0143 & 0.0145\\
0.50 & 0.174 & 0.3775 & 0.0818 & 0.0823 & 0.2601 & 0.0152 & 0.0169 & 0.0865 & 0.0146 & 0.0148\\
0.60 & 0.127 & 0.2595 & 0.0507 & 0.0519 & 0.2757 & 0.0114 & 0.0139 & 0.0622 & 0.0105 & 0.0108\\
0.60 & 0.133 & 0.2799 & 0.0523 & 0.0535 & 0.2799 & 0.0116 & 0.0141 & 0.0675 & 0.0109 & 0.0112\\
0.60 & 0.153 & 0.1833 & 0.0856 & 0.0860 & 0.2660 & 0.0207 & 0.0221 & 0.0469 & 0.0196 & 0.0197\\
0.60 & 0.161 & 0.2449 & 0.0982 & 0.0985 & 0.2705 & 0.0222 & 0.0236 & 0.0613 & 0.0211 & 0.0212\\
0.60 & 0.171 & 0.3730 & 0.1223 & 0.1226 & 0.2783 & 0.0240 & 0.0252 & 0.0885 & 0.0232 & 0.0233\\
0.70 & 0.152 & 0.1894 & 0.0567 & 0.0577 & 0.2774 & 0.0138 & 0.0160 & 0.0493 & 0.0134 & 0.0137\\
0.70 & 0.158 & 0.3134 & 0.0526 & 0.0536 & 0.2857 & 0.0115 & 0.0141 & 0.0768 & 0.0107 & 0.0110\\
0.70 & 0.166 & 0.2714 & 0.0545 & 0.0551 & 0.2803 & 0.0124 & 0.01478 & 0.0681 & 0.0116 & 0.0118\\
0.70 & 0.205 & 0.4366 & 0.0882 & 0.0893 & 0.2804 & 0.0160 & 0.0179 & 0.1032 & 0.0169 & 0.0172\\
0.70 & 0.214 & 0.4480 & 0.2080 & 0.2082 & 0.2734 & 0.0366 & 0.0374 & 0.1041 & 0.0367 & 0.0368\\
0.70 & 0.228 & 0.2756 & 0.0799 & 0.0803 & 0.2688 & 0.0175 & 0.0191 & 0.0733 & 0.0178 & 0.0180\\
0.70 & 0.236 & 0.3779 & 0.0753 & 0.0758 & 0.2712 & 0.0144 & 0.0163 & 0.0952 & 0.0153 & 0.0155\\
0.70 & 0.283 & 0.3729 & 0.1029 & 0.1032 & 0.2513 & 0.0186 & 0.0199 & 0.0959 & 0.0205 & 0.0207\\
0.80 & 0.185 & 0.3167 & 0.0726 & 0.0732 & 0.2851 & 0.0156 & 0.01759 & 0.0789 & 0.0152 & 0.0154\\
0.80 & 0.237 & 0.3768 & 0.0625 & 0.0634 & 0.2768 & 0.0125 & 0.0147 & 0.0946 & 0.0123 & 0.0127\\
0.80 & 0.245 & 0.3508 & 0.0606 & 0.0613 & 0.2666 & 0.0119 & 0.0141 & 0.0875 & 0.0121 & 0.0124\\
0.80 & 0.252 & 0.4501 & 0.1384 & 0.1387 & 0.2663 & 0.0239 & 0.0251 & 0.1060 & 0.0241 & 0.0242\\
0.80 & 0.261 & 0.2945 & 0.1191 & 0.1193 & 0.2605 & 0.0249 & 0.0260 & 0.0770 & 0.0253 & 0.0254\\
0.80 & 0.269 & 0.3988 & 0.0582 & 0.0589 & 0.2691 & 0.0108 & 0.0132 & 0.1010 & 0.0115 & 0.0118\\
0.80 & 0.279 & 0.3388 & 0.0591 & 0.0596 & 0.2716 & 0.0119 & 0.0142 & 0.0922 & 0.0130 & 0.0132\\
0.80 & 0.289 & 0.2823 & 0.0590 & 0.0594 & 0.2642 & 0.0126 & 0.0146 & 0.0795 & 0.0138 & 0.0140\\
0.80 & 0.353 & 0.2637 & 0.0888 & 0.0890 & 0.2327 & 0.0174 & 0.0186 & 0.0751 & 0.0208 & 0.0209\\
0.80 & 0.369 & 0.2732 & 0.1002 & 0.1004 & 0.2256 & 0.0188 & 0.0198 & 0.0774 & 0.0233 & 0.0234\\
0.90 & 0.236 & 0.4082 & 0.0944 & 0.0949 & 0.2842 & 0.0181 & 0.0198 & 0.1003 & 0.0183 & 0.0185\\
0.90 & 0.325 & 0.2924 & 0.0627 & 0.0630 & 0.2421 & 0.0121 & 0.0139 & 0.0774 & 0.0137 & 0.0138\\
0.90 & 0.337 & 0.1786 & 0.0571 & 0.0573 & 0.2232 & 0.0120 & 0.0136 & 0.0489 & 0.0138 & 0.0139\\
0.90 & 0.350 & 0.2688 & 0.0616 & 0.0619 & 0.2200 & 0.0113 & 0.0129 & 0.0690 & 0.0131 & 0.0132\\
0.90 & 0.435 & 0.3300 & 0.1162 & 0.1164 & 0.1636 & 0.0147 & 0.0154 & 0.0706 & 0.0194 & 0.0195\\
0.90 & 0.457 & 0.3223 & 0.1259 & 0.1260 & 0.1430 & 0.0141 & 0.0147 & 0.0633 & 0.0194 & 0.0194\\
0.90 & 0.481 & 0.0950 & 0.0952 & 0.0952 & 0.1357 & 0.0145 & 0.0150 & 0.0224 & 0.0208 & 0.0208\\
0.90 & 0.509 & 0.3363 & 0.1631 & 0.1632 & 0.1469 & 0.0183 & 0.0188 & 0.0744 & 0.0281 & 0.0282\\
1.00 & 0.326 & 0.3221 & 0.0753 & 0.0756 & 0.2569 & 0.0148 & 0.0165 & 0.0860 & 0.0163 & 0.0164\\
1.00 & 0.337 & 0.3511 & 0.0922 & 0.0926 & 0.2507 & 0.0171 & 0.0184 & 0.0912 & 0.0191 & 0.0193\\
1.00 & 0.389 & 0.0749 & 0.0844 & 0.0845 & 0.1967 & 0.0179 & 0.0188 & 0.0210 & 0.0225 & 0.0226\\
1.00 & 0.405 & 0.2822 & 0.1031 & 0.1033 & 0.2101 & 0.0172 & 0.0182 & 0.0729 & 0.0224 & 0.0225\\
1.00 & 0.565 & 0.0923 & 0.0978 & 0.09778 & 0.1432 & 0.0158 & 0.0164 & 0.0257 & 0.0252 & 0.0252\\\hline
\end{tabular}
\end{center}
\caption{Deuterium structure functions $F_2$, $R$, and $F_L$ extracted using the Rosebluth separation technique.}
\end{table*} 

\begin{figure}[t]
 \begin{center}
\includegraphics[scale=0.4]{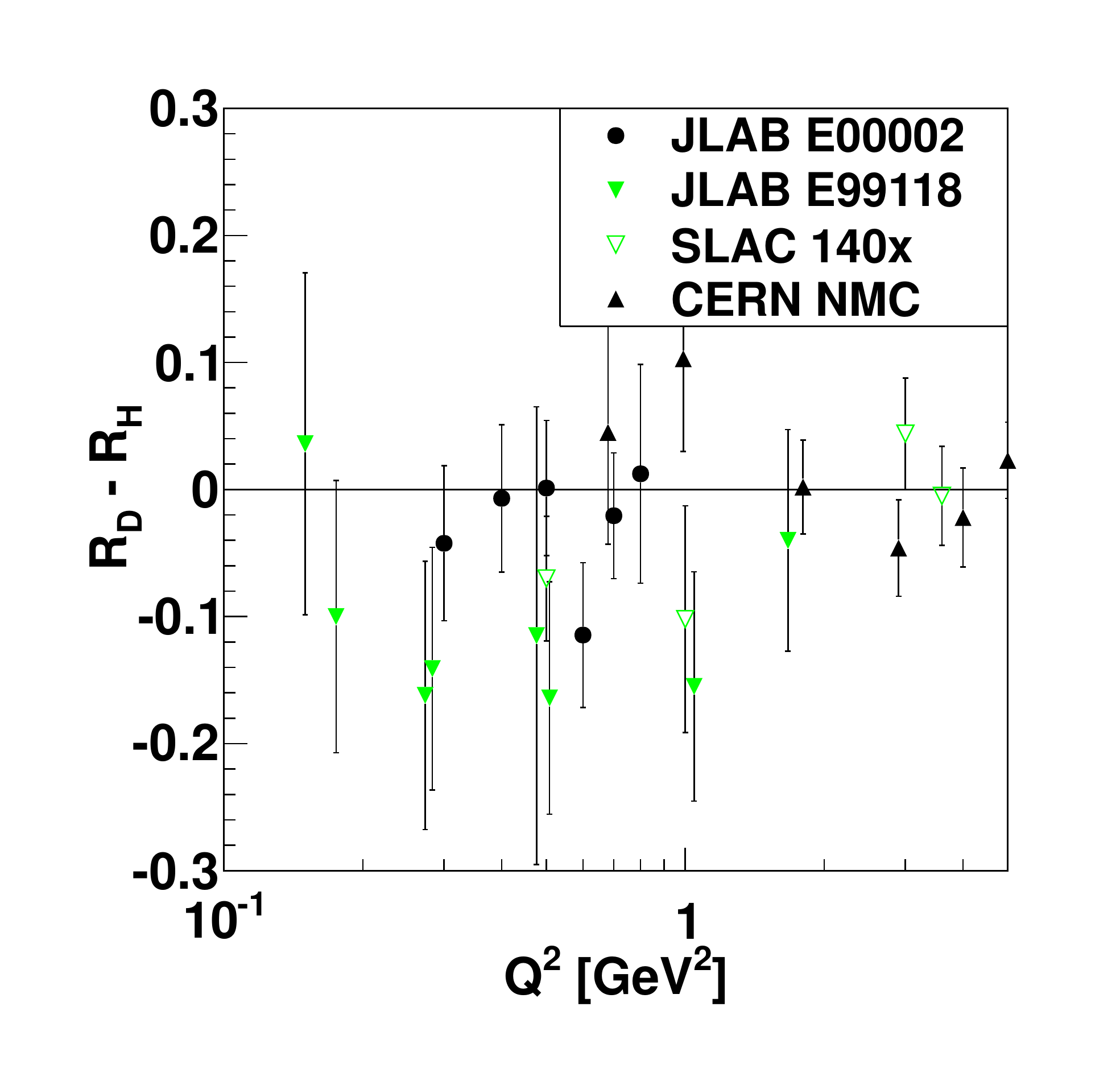}
\end{center}
  \caption{ (Color online) Difference of $R_H$  and $R_D$ as a function of $Q^2$. }
\label{fig:fig4}
\end{figure}

We show in Fig.\ \ref{fig:fig4} the deuteron-proton difference, $R_D - R_H$, as a function of $Q^2$. The data are a weighted average over $x$ of data beyond the nucleon resonance region, $W >$ 2 GeV.
There is a systematic overall shift of the data towards negative values, indicating that below $Q^2=1.5$~GeV$^2$ $R$ for the deuteron might be smaller than that for the proton.
For this kinematic range a weighted average of all available data, including the new E00--002 data, yields $R_D - R_H = -0.042\pm 0.018$ for $Q^2< 5$~GeV$^2$ , in agreement with the result obtained by a similar global average presented in\ \cite{vladas}, which did not include the present data and obtained $R_D - R_H = -0.054\pm 0.029$. While this difference is a small quantity, it should be noted that the value of $R$ in this region is only on average 0.1. Hence, this is a $\geq 30$\% effect.

It is useful to note that there may be some $x$ dependence to these data as well. It would be reasonable, for instance, for any longitudinal EMC effect in deuterium to have some $x$ dependence in this kinematic regime given that the cross section has a well--known $x$--dependence.

Although there is nothing to prevent it, there is no theoretical model that
predicts the neutron structure function ratio $R_n$ to differ from the proton $R_H$.
Therefore, in most previous analyses, $R_D$ was assumed to equal $R_H$, thus
neglecting possible nuclear binding or EMC--effect type phenomena in the deuteron -- or
an inherent difference between $R_n$ and $R_H$. Previous measurements had, moreover,
generally confirmed no difference, i.e. $R_D = R_H$\ \cite{arneodo,140X} for $Q^2\geq 1.5$~GeV$^2$. However, in the high momentum domain of these measurements, $R_H \approx 0$ and any difference is practically immeasurable at large $Q^2$ values. This fact may resolve the apparent discrepancy between the higher $Q^2$ results and the results shown here.

\begin{figure}
 \begin{center}
\includegraphics[scale=0.4]{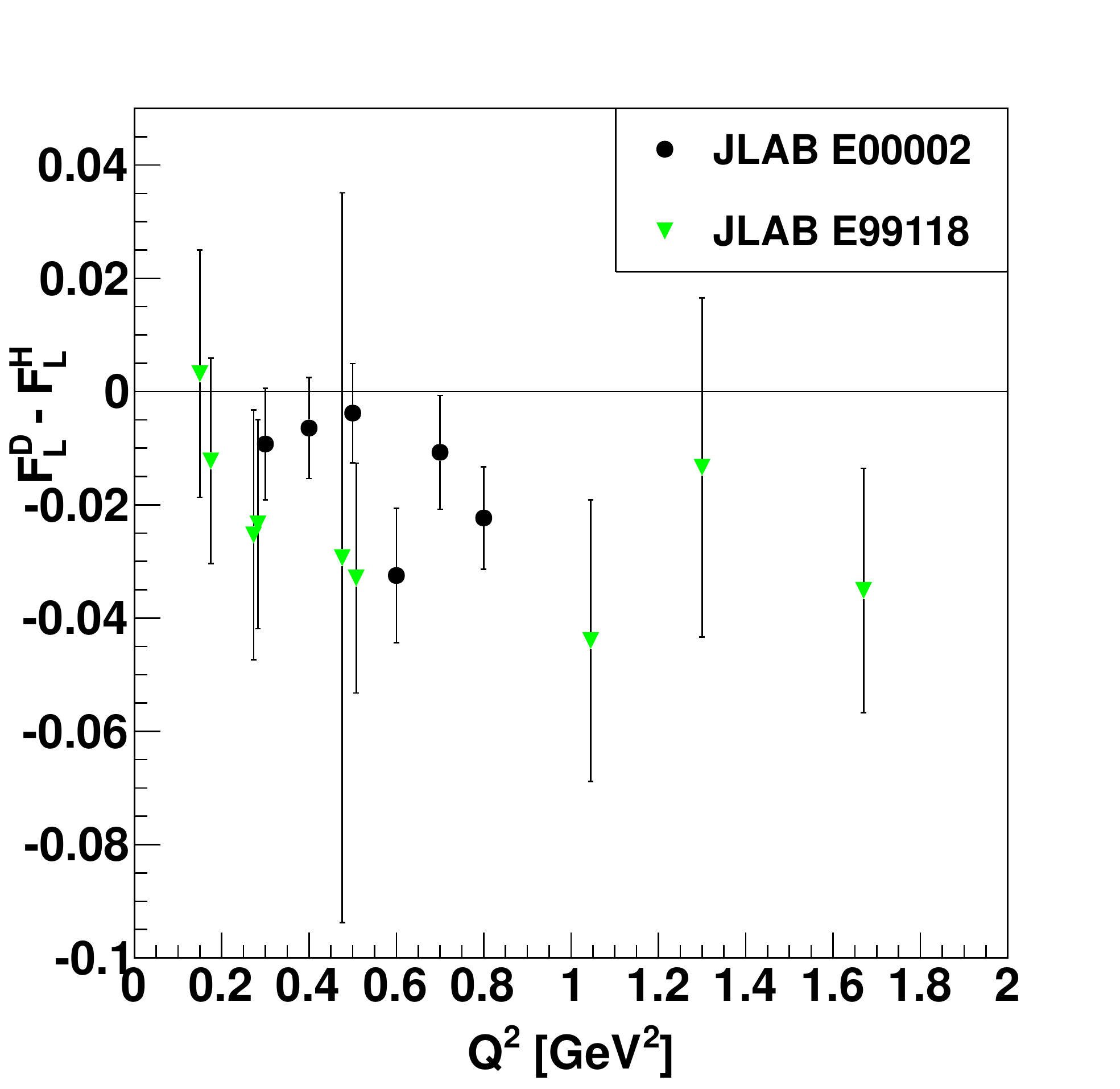}
\end{center}
  \caption{ (Color online) Difference of $F_L^H$  and $F_L^D$ as a function of $Q^2$. }
\label{fig:fig5}
\end{figure}

In the framework of the leading-twist formalism, global QCD fits
to available nuclear data allow for enhancements of valence-quark and gluon distributions
in nuclei as compared to a free proton\ \cite{Accardi01,guzey}.
The $F_L$ structure function  would be uniquely sensitive to such an enhanced gluon distribution\ \cite{guzey}. To investigate this possibility we plot in Fig.\ \ref{fig:fig5} the structure function $F_L^D - F_L^H$ as a function of $Q^2$. The plot shows data from Jefferson Lab experiments: E99--118 (inverted triangles) and E00--002 (closed circles). The SLAC and CERN experiments shown in Fig.\ \ref{fig:fig4} do not extract the longitudinal structure function $F_L$ which is purely a function of the longitudinal cross section, only the ratio $R=\sigma_L/\sigma_T$. The low--mass deuteron may not be the optimum target for studying gluonic nuclear effects, however some nuclear dependence to $F_L$, or perhaps $F_L^n$ differing from $F_L^H$, is clearly observable in the longitudinal channel at the kinematics accessible here.

\section{CONCLUSION}

In conclusion, we report new separated structure function measurements
for hydrogen and deuterium at low $Q^2$ (below 2~GeV$^2$).
Available parameterizations and theoretical predictions for the longitudinal structure-function $F_L$ agree reasonably well when extrapolated to the low $Q^2$ (as low as $Q^2=1.0$~GeV$^2$)
and large $x$ ($0.1\leq x \leq 0.6$) kinematics of these new data. Remarkably, the global data
set seems to smoothly converge toward zero below $Q^2 = 1.0$~GeV$^2$,
even while it is still far above the current conservation limit.
Additionally, the deuterium data seem to confirm a smaller ratio $R = \sigma_L/\sigma_T$
than the hydrogen data and therefore indicate either a nuclear dependence to the longitudinal response of the nucleons in the deuteron, and hence a nuclear dependence to the $F_L$ structure function, or a difference between the neutron and proton $F_L$ structure functions.

This work was supported by the United States Department of Energy (DOE)
Contract No.~DE-AC05-06OR23177, under which Jefferson Science Associates,
LLC operates Jefferson Lab.  
This work was supported in part by the U.S. Department of Energy under Grant No. DE--AC02--06CH11357
 and the National Science Foundation under Grants No. PHY--0245045, PHY--0555510, PHY--0653440, and PHY--1002644.

\bibliography{}

%

\end{document}